\documentclass[sigconf]{acmart}
\acmDOI{}
\acmISBN{}
\AtBeginDocument{%
  }

\usepackage{multirow}
\usepackage{booktabs}
\usepackage{subcaption}
\usepackage{graphicx}
\usepackage{soul}
\setcopyright{rightsretained}
\usepackage[ruled,vlined]{algorithm2e}
\acmConference[CARS @ RecSys '26]
  {T
  the 20th ACM Conference on Recommender Systems}
  {September 28--October 2, 2026}
  {Minneapolis, MN, USA}

\acmBooktitle{The 20th ACM Conference on Recommender Systems (RecSys '26), September 28--October 2, 2026, Minneapolis, MN, USA}

\begin{document}

\title[Gated Multimodal Fusion in Sequential Recommendation]{Deciding When to Rely on Visual Information: Gated Multimodal Fusion in Sequential Recommendation}



\author{Natalija Glisovic}
\email{natalija.glisovic@ingka.ikea.com}
\orcid{1234-5678-9012}
\authornotemark[1]
\affiliation{%
  \institution{IKEA Retail (Ingka Group) and KTH Royal Institute of Technology}
 \city{Malmö}
  \country{Sweden}
}

\author{Danica Kragic}
\email{dani@kth.se}
\affiliation{%
  \institution{KTH Royal Institute of Technology}
  \city{Stockholm}
  \country{Sweden}}
  
\author{Martin Tegner}
\email{martin.tegner@ingka.ikea.com}
\affiliation{%
  \institution{IKEA Retail (Ingka Group)}
  \city{Malmö}
  \country{Sweden}}

\renewcommand{\shortauthors}{Glisovic et al.}

\begin{abstract}
Multimodal sequential recommender systems commonly fuse visual and collaborative signals uniformly, treating visual features as generically informative regardless of item or user context. We argue that visual utility, defined as the contribution of visual signals to recommendation quality, is a {latent} contextual {variable} that depends on both the item and the user's interaction history rather than a fixed item property. To model this variability, we introduce VisGate, a framework that makes adaptive item-level fusion decisions conditioned on item embeddings and the user's current sequence context. Visual representations are learned through a contrastive objective over sequential co-occurrence patterns, preserving complementarity with collaborative embeddings rather than aligning them into a shared space. Beyond achieving competitive recommendation performance, VisGate's learned gate serves as a measurement tool for understanding when and why visual information is beneficial. Our analyses show that visual utility varies across items, increases under interaction sparsity when collaborative signals are weak, and correlates with visual distinctiveness in semantically meaningful ways. Together, these findings highlight the importance of both fine-grained fusion and modality complementarity, while demonstrating that item-level visual utility can be estimated and interpreted through learned gating behaviour.
\end{abstract}





\maketitle
\section{Introduction}\label{sec:intro}
Context-aware recommender systems have long recognised that when, where, and how a user interacts with an item shapes what constitutes a relevant recommendation \cite{adomavicius2011cars, mateos2025carsreview}. However, most systems treat contextual signals as predefined and static. We argue that which modality to rely on is itself a contextual decision that depends not on fixed item properties, but on the user's interaction state at the point of recommendation.

\begin{figure}[h!]
\centering
\includegraphics[trim=7mm 8mm 0mm 0mm, clip, 
width=0.5\textwidth]{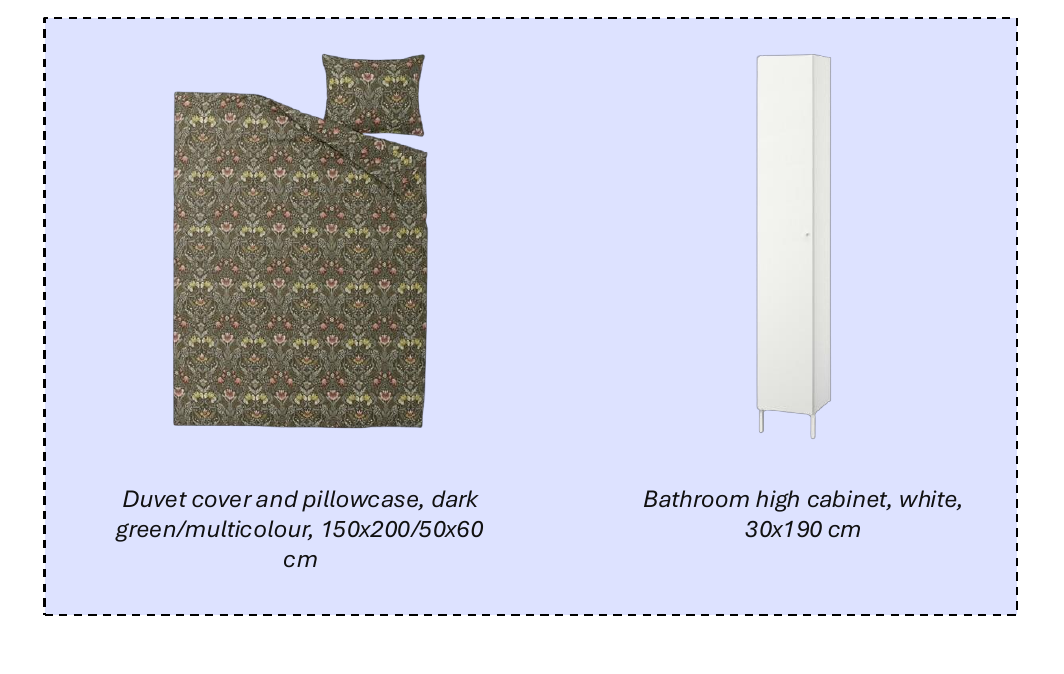}
\caption{Examples from the IKEA dataset where visual features 
provide richer information than text descriptions.}
\label{fig:example_text}
\end{figure}

Sequential recommenders capture temporal dynamics of user behaviour through interaction patterns, but face well-known limitations under data sparsity. Learned ID-based embeddings struggle to represent cold-start items or users with limited history \cite{kang2028sasrec}. Incorporating item content such as images has shown promise in addressing these limitations \cite{malitesta2024missing}, giving rise to multimodal sequential recommendation \cite{hou2022unisrec, hu2023mmsr}. Yet, a question that remains less explored is when visual content helps. Product images are often more available than text in e-commerce, where descriptions are frequently sparse, generic, or missing. Figure \ref{fig:example_text} illustrates this with IKEA examples where a patterned duvet is described only as "green/multicolour" and a cabinet simply as "white", text that fails to capture the visual characteristics most likely to influence purchase decisions \cite{yin2025securing}. Most existing approaches fuse modalities uniformly, treating visual features as generically informative regardless of item or context \cite{pomo2025multimodal}, yet naive fusion introduces noise when modalities carry redundant or conflicting signals \cite{malitesta2024missing, bruun2024missingmodal}.

We introduce VisGate, a framework that treats modality utility, i.e. the item-level contribution of a given modality to recommendation quality, as a latent contextual variable inferred from interaction patterns and user state. A context-aware adaptive gating mechanism learns when to rely on visual versus collaborative signals, conditioned on both item-level embeddings and the user's current sequence representation. Visual representations are learned via a contrastive objective over sequential co-occurrence patterns, preserving complementarity with collaborative embeddings rather than aligning them into a shared space. This design is motivated by three research questions:

\begin{itemize}
    \item[\textbf{RQ1:}] Is visual utility item-dependent, or can it be captured by uniform fusion across all items?
    \item[\textbf{RQ2:}] Does visual utility increase under interaction sparsity, when collaborative signals are weakest?
    \item[\textbf{RQ3:}] Does visual utility correlate with visual distinctiveness, such that the gate assigns higher visual weight to items whose appearance is more informative?
\end{itemize}

We validate VisGate on four datasets, using the learned gate as a measurement tool to answer these questions. Our main contributions are:

\begin{itemize}
    \item We introduce modality utility as a concept and show that it is a latent, dynamic contextual variable shaped by interaction history and user state, not a fixed item property.
    \item We propose VisGate, combining co-occurrence grounded visual projection with an item-level adaptive gating mechanism that conditions fusion decisions on the user's current sequence context.
    \item We validate three hypotheses about when visual signals help, providing empirical evidence that fusion granularity and complementarity preservation matter as much as the choice of fusion mechanism.
    \item We analyse failure cases where the gate misleads, identifying a practical interaction-count threshold as a post-hoc correction.
\end{itemize}

\section{Related Work}
Sequential recommendation has evolved from RNN-based approaches such as GRU4Rec \cite{hidasi2016gru4rec} to transformer architectures including SASRec \cite{kang2028sasrec} and BERT4Rec \cite{sun2019bert4rec}. Despite strong performance, purely ID-based models struggle with data sparsity and cold-start items, motivating the incorporation of item content.

Multimodal recommender systems address this by augmenting ID-based embeddings with visual and textual signals \cite{oramas2017deep}. Classical visual models such as VBPR \cite{he2016vbpr} and DVBPR \cite{kang2017visually} established foundations for image-based recommendation, while graph-based methods such as MMGCN \cite{wei2019mmgcn} and FREEDOM \cite{zhou2023freedom} propagate multimodal signals through interaction graphs. Recent work has questioned whether multimodal gains stem from cross-modal understanding or increased model complexity \cite{pomo2025multimodal}, and shown that naive handling of missing modalities can harm performance \cite{malitesta2024missing}, highlighting the need for more principled fusion strategies.

A directly relevant thread concerns how contextual information is modelled. Early work on context-aware recommendations distinguishes a representational view, where context is a set of predefined dimensions such as location, from an interactional view, where context emerges dynamically and is not fully observable in advance \cite{adomavicius2011cars, adomavicius2022cars}. More recent work has pushed toward latent context. \cite{beutel2018latentcross} showed that fusing contextual embeddings with an RNN's hidden state via element-wise products outperforms naive feature concatenation, while \cite{chang2023latentuserintent} formalised user intent as a latent variable inferred from interaction sequences using variational autoencoders, without requiring explicit context labels.

In multimodal sequential recommendation, early work such as UniSRec \cite{hou2022unisrec} and TedRec \cite{xu2024tedrec} replaced or augmented item IDs with text-based representations to enable transferable sequence models. Subsequent methods shifted toward adaptive fusion. MMSR \cite{hu2023mmsr}, HM4SR \cite{zhang2025hm4sr}, and CAMMSR \cite{xu2026cammsr} introduce gating or mixture-of-experts mechanisms to route across modalities at the sequence or step level, while MP4SR \cite{zhang2024mp4sr} uses contrastive pre-training over mixed-modality sequences to regularise representations. Together, these methods confirm that static, uniform fusion introduces noise when modalities carry redundant or conflicting signals.

VisGate situates itself within the latent context modelling tradition, treating visual utility, i.e. the item-level contribution of visual signals to recommendation quality, as a latent contextual variable inferred from interaction patterns and the user's current sequence state. Unlike prior multimodal sequential methods that operate at sequence or step level and align modalities into shared spaces, VisGate makes fusion decisions at the item level and deliberately preserves complementarity between visual and collaborative representations, enabling the interpretable gating behaviour we analyse in Section \ref{sec:results}. 

\section{Methodology}
\begin{figure*}[t]
    \centering
    \includegraphics[trim=0 255pt 0 0, clip, width=\textwidth]{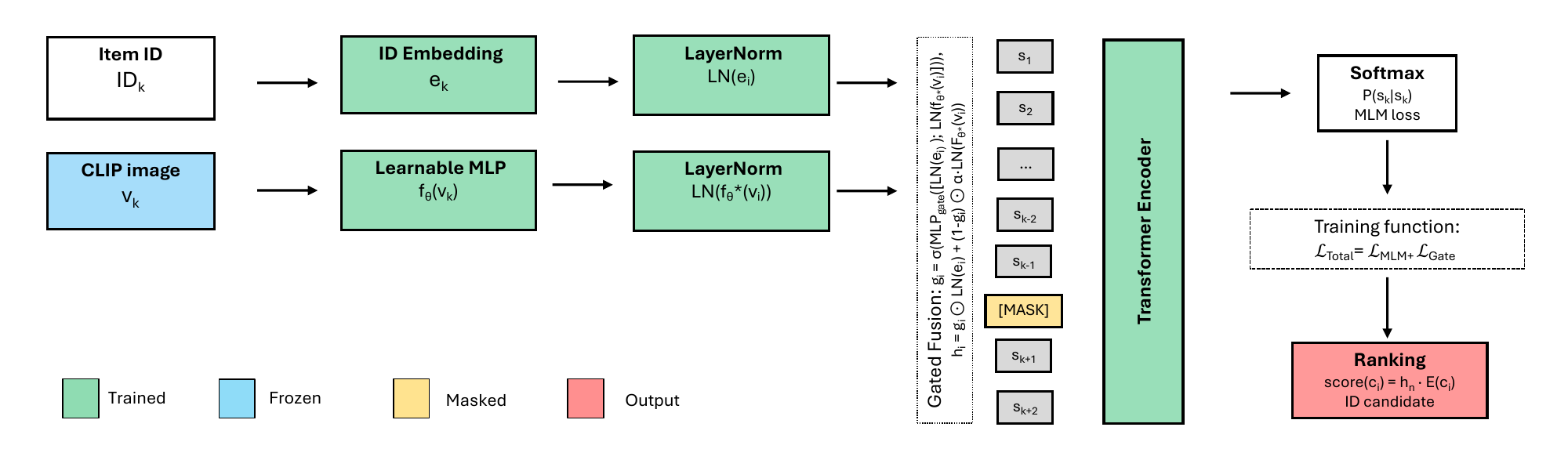}
    \caption{An overview of VisGate architecture}
    \label{fig:method_overview}
\end{figure*}

This section introduces VisGate, which serves both as a competitive recommender and as a tool for studying context-dependent modality utility. We describe the three components: co-occurrence grounded visual projection, context-aware adaptive gating, and the training procedure. The overall architecture of VisGate is illustrated in Figure \ref{fig:method_overview}.

\subsection{Co-occurrence Grounded Visual Projection}
To learn visual representations that capture user-relevant patterns, we employ a projection supervised by sequential co-occurrence patterns. The key insight is that items appearing in temporal proximity in user sequences often share visual characteristics that influence user preferences, providing a natural source of supervision for visual feature learning. Crucially, this projection maps visual features into a co-occurrence grounded space rather than into the collaborative embedding space itself, preserving complementarity between the two modalities.

We train a multilayer perceptron (MLP) projector $f_\theta : \mathbb{R}^d \rightarrow \mathbb{R}^d$ that maps visual features from the pre-trained vision encoder to a co-occurrence grounded 
space:
\begin{equation}
f_\theta(v_i) = \text{Linear}_2(\text{GELU}(\text{Linear}_1(\phi(v_i))))
\end{equation}
where $\text{Linear}_1 : \mathbb{R}^d \rightarrow \mathbb{R}^{h}$ and $\text{Linear}_2 : \mathbb{R}^{h} \rightarrow \mathbb{R}^d$ are linear 
transformations with hidden dimension $h$. Unlike approaches that project 
visual features into collaborative embedding space~\cite{he2016vbpr}, this 
supervision keeps visual representations in a complementary space.

We optimize the projector using a contrastive objective over positive item pairs extracted from user sequences based on temporal proximity:
\begin{equation}
\mathcal{P} = \{(i, j) \mid \exists u, t_1, t_2 : s^u_{t_1} = i, 
s^u_{t_2} = j, |t_1 - t_2| \leq w\}
\end{equation}
where $w$ is a temporal window size. The contrastive loss encourages 
visual features of temporally co-occurring items to be similar:
\begin{equation}
\mathcal{L}_{\text{co-occ}} = -\frac{1}{|\mathcal{B}|} \sum_{(i,p) \in 
\mathcal{B}} \log \frac{\exp(\text{sim}(f_\theta(v_i), f_\theta(v_p))/\tau)}
{\sum_{j \in \mathcal{N}(i)} \exp(\text{sim}(f_\theta(v_i), 
f_\theta(v_j))/\tau)}
\end{equation}
where $\mathcal{B} \subset \mathcal{P}$ is a batch of positive pairs, $\text{sim}(\cdot, \cdot)$ denotes cosine similarity, $\mathcal{N}(i)$ are sampled negative items, and $\tau$ is a learnable temperature parameter. Co-occurring pairs exhibit significantly higher visual similarity than random pairs (Mann--Whitney $U$, $p < 10^{-10}$, effect size $r \approx 0.2$--$0.3$), confirming that co-occurrence 
is a meaningful, if noisy, supervision signal.

\subsection{Context-Aware Adaptive Gating}
After pre-training, we introduce a context-aware adaptive gating mechanism that learns to balance ID and visual representations for each item. The key difference from prior gated fusion approaches \cite{hu2023mmsr, xu2026cammsr} is that our gate is conditioned on both item-level embeddings and the user's current sequence context. This reflects the observation that visual utility is not a fixed item property. The same item may benefit more from visual signals early in a user's interaction history, when collaborative embeddings are sparse, than later when rich interaction patterns have accumulated. By conditioning on the transformer's hidden state at the masked position, the gate can adapt its fusion decision to the user's context at the point of recommendation.

Concretely, let $\mathbf{z} \in \mathbb{R}^d$ denote the transformer's final hidden state at the masked position, capturing the user's current sequence context. The gating function takes layer-normalized embeddings from both modalities alongside this contextual signal:
\begin{equation}
g_i = \sigma(\text{MLP}_{\text{gate}}([\text{LN}(e_i);\, 
\text{LN}(f_{\theta^*}(v_i));\, \text{LN}(\mathbf{z})]))
\end{equation}
\begin{equation}
h_i = g_i \odot \text{LN}(e_i) + (1 - g_i) \odot 
(\alpha \cdot \text{LN}(f_{\theta^*}(v_i)))
\end{equation}
where $[\cdot;\cdot]$ denotes concatenation, $\text{LN}(\cdot)$ represents layer normalization, $\theta^*$ are the pre-trained projector parameters, $\alpha \in \mathbb{R}$ is a learnable scalar that scales the visual component, and $\text{MLP}_{\text{gate}}$ is a two-layer network with ReLU activation and dropout. Layer normalization ensures both modalities and the sequence context have comparable scales, so the gate makes decisions based on semantic relevance rather than magnitude differences. The gate $g_i \in [0, 1]^d$ produces element-wise weights determining the contribution of collaborative versus visual signals to the final item representation $h_i$.

\subsection{Training}
The final training objective combines the standard masked language modeling (MLM) loss from BERT4Rec with gating regularization terms:
\[
\mathcal{L}_{\text{total}} = \mathcal{L}_{\text{MLM}} + \mathcal{L}_{\text{gate}}
\]
\[
\mathcal{L}_{\text{MLM}} = -\sum_{i \in \mathcal{M}} \log P(s_i \mid h_{<i})
\]
where $\mathcal{M}$ denotes the set of masked positions, and $P(s_i \mid h_{<i}) = \text{softmax}(\mathbf{z}_i^{\top} \mathbf{H})$ is computed using the transformer's final hidden state $\mathbf{z}_i$ and the gated item-embedding matrix $\mathbf{H} \in \mathbb{R}^{I \times d}$.

During training, we jointly optimize the transformer parameters, ID embeddings, gating mechanism, and scaling factor, while keeping the pre-trained projector $\theta^*$ fixed:
\[
\boldsymbol{\Phi}^*, \mathbf{E}^*, \boldsymbol{\psi}^*, \alpha^* = 
\arg\min_{\boldsymbol{\Phi}, \mathbf{E}, \boldsymbol{\psi}, \alpha} 
\mathcal{L}_{\text{total}}
\]

To encourage meaningful gating adaptation, we incorporate auxiliary losses:
\[
\mathcal{L}_{\text{gate}} = \lambda_1 \mathcal{L}_{\text{sparsity}} + 
\lambda_2 \mathcal{L}_{\text{util}}
\]
where $\mathcal{L}_{\text{sparsity}} = \mathbb{E}_{i}[4g_i(1 - g_i)]$ encourages decisive gating decisions, and $\mathcal{L}_{\text{util}} = |\mathbb{E}_{i}[g_i] - 0.5|$ prevents collapse to a single modality by encouraging balanced utilization across the item set. We apply higher learning rates for the gating mechanism to enable quick adaptation, standard rates for ID embeddings, and keep the visual projector frozen.

\section{Experiments}
In this section, we present the experimental evaluation of VisGate. We begin by describing the experimental setup, including the datasets, baseline models, and evaluation settings. Next, we present the main result, followed by targeted analyses addressing the research questions introduced in Section \ref{sec:intro}.

\subsection{Experimental Setup}
\subsubsection{Datasets}
We evaluate VisGate on three public Amazon Review datasets \cite{ni2019amazon}: Amazon Scientific, Amazon Video Games, and Amazon Health \& Household. Each dataset contains user–item interactions accompanied by item images. For each user, we order interactions based on timestamps to construct sequential recommendation data. In addition to the Amazon datasets, we include a proprietary IKEA dataset with a similar schema. 

Following prior research \cite{kim2025image, qin2025more, wei2024towards}, we filter out users and items with fewer than five interactions. The exception is the Amazon Scientific dataset, which we retain as a sparse benchmark to evaluate model robustness under limited interaction conditions. Here, we only remove users and items with fewer than two interactions. Table \ref{tab:datasets} summarizes the statistics of the datasets. 

\begin{table}[t]
\centering
\caption{Summary statistics of the datasets used in experiments. Avg. Len is the average sequence length.}
\label{tab:datasets}
\begin{tabular}{lccc}
\toprule
\textbf{Dataset} & \textbf{\#Users} & \textbf{\#Items} & \textbf{Avg. Len} \\
\midrule
IKEA & 49,962 & 23,551 & 7.15 \\
Amazon Scientific & 38,861 & 22,909 & 4.05 \\
Amazon Video Games & 71,473 & 17,934 & 4.49 \\
Amazon Health \& Household & 159,404 & 76,806 & 8.52 \\
\bottomrule
\end{tabular}
\end{table}

\subsubsection{Baselines}
To evaluate the effectiveness of VisGate, we compare against four categories of recommender systems: Collaborative Filtering (CF): BERT4Rec \cite{sun2019bert4rec}, SASRec \cite{kang2028sasrec}, GRU4Rec \cite{hidasi2016gru4rec}. Visually-aware recommender systems: VBPR \cite{he2016vbpr}, DVBPR \cite{kang2017visually}, AMR \cite{tang2020adversarial}. Multimodal recommender systems:  MMGCN \cite{wei2019mmgcn}, FREEDOM \cite{zhou2023freedom}. Lastly, multimodal sequential recommender systems: UniSRec \cite{hou2022unisrec}, MP4SR \cite{zhang2024mp4sr}, MMSR \cite{hu2023mmsr}, HM4SR \cite{zhang2025hm4sr}. 

We implement all baselines using their original GitHub repositories, or the standardized implementations from MMRec\footnote{\url{https://github.com/enoche/MMRec}}.

\subsubsection{Evaluation Settings}
We follow the leave-one-out evaluation protocol \cite{hebert2025flare, zhou2022filter,  kim2025image}. Specifically, for each user sequence, we reserve the last interacted item as the test instance and the second-to-last item for validation, using all preceding interactions for model training. To assess statistical reliability, we average all results over 5 independent runs with different random seeds and mark improvements as significant where a paired t-test against the strongest baseline on each dataset yields p < 0.01.

For evaluation, we use Hit Ratio (Hit@$k$) and Normalized Discounted Cumulative Gain (NDCG@$k$) with $k=5,10$, where Hit@$k$ checks if the ground-truth item appears in the top-$k$ recommendations, and NDCG@$k$ additionally accounts for its ranking position \cite{kim2025image}.

\subsubsection{Implementation Details}
We implement VisGate in PyTorch and train all models using Adam, with gradient norms clipped to $1.0$ and early stopping based on validation Hit@10 (patience of 5 epochs, with a maximum budget of 50 epochs). We retain the checkpoint achieving the best validation performance. The gating MLP uses twice the base learning rate and a weight decay of $10^{-5}$ to enable faster adaptation of the fusion mechanism, while all remaining parameters use the base learning rate with a weight decay of $10^{-4}$.

Capacity-related hyperparameters are fixed and matched across VisGate and all baselines: hidden dimension of $256$, $2$ Transformer layers, $4$ attention heads, maximum sequence length of $50$, batch size of $128$, dropout rate of $0.1$, and masking probability of $0.15$. The remaining VisGate-specific hyperparameters, which have no direct counterparts in the baselines, are selected independently for each dataset via random search on the validation split. Specifically, the learning rate is selected from ${0.0001, 0.0005, 0.001, 0.005, 0.01}$, $\lambda_1$ and $\lambda_2$ are sampled log-uniformly from $[0.001, 0.05]$, and the co-occurrence window $w$ is selected from ${1,2,3,5,7}$. The visual projector $f_{\theta}$ is pre-trained separately using the contrastive objective and subsequently frozen. This separation ensures that the gate's fusion decisions can be attributed to the modality-combination mechanism rather than to a visual projector that continues to reshape the visual representation during training. 

\subsection{Results}\label{sec:results}
Table \ref{tab:main_results} presents results across all four datasets. VisGate achieves the strongest overall performance. The largest gains are on the IKEA and Amazon Scientific, which are the two sparsest datasets. On denser datasets gains are more modest. In these cases, methods that incorporate text alongside images, such as FREEDOM and HM4SR, are more competitive. We organise the analysis around our three research questions.

\begin{table*}[t]
\centering
\scriptsize
\caption{Performance comparison across datasets. Best result is bolded, second best is underlined. \% Improve shows relative improvement of VisGate over the best baseline. $\dagger$ denotes significant improvement over the best baseline.}
\setlength{\tabcolsep}{2.5pt}
\begin{tabular}{l l | cccc | cccc | cccc | cccc}
\toprule
& & \multicolumn{4}{c|}{IKEA} & \multicolumn{4}{c|}{Amazon Scientific} & \multicolumn{4}{c|}{Amazon Video Games} & \multicolumn{4}{c}{Amazon Health \& Household} \\
\cmidrule(r){3-6} \cmidrule(r){7-10} \cmidrule(r){11-14} \cmidrule(r){15-18}
Category & Model & H@5 & N@5 & H@10 & N@10 & H@5 & N@5 & H@10 & N@10 & H@5 & N@5 & H@10 & N@10 & H@5 & N@5 & H@10 & N@10 \\
\midrule
\multirow{3}{*}{CF}
& BERT4Rec & 0.1352 & 0.1290 & 0.1881 & 0.1469 & 0.0643 & 0.0448 & 0.1163 & 0.0620 & 0.1332 & 0.1230 & 0.1754 & 0.1345 & 0.1045 & 0.0853 & 0.1517 & 0.1004 \\
& SASRec   & 0.1247 & 0.1058 & 0.1782 & 0.1229 & 0.0817 & 0.0574 & 0.1292 & 0.0761 & 0.1229 & 0.1210 & 0.1702 & 0.1239 & 0.1041 & 0.0766 & 0.1339 & 0.0934 \\
& GRU4Rec  & 0.1139 & 0.1002 & 0.1699 & 0.1189 & 0.0640 & 0.0429 & 0.1098 & 0.0622 & 0.1328 & 0.1203 & 0.1744 & 0.1341 & 0.1053 & 0.0799 & 0.1347 & 0.0931 \\
\midrule
\multirow{3}{*}{Visually-aware}
& VBPR     & 0.1655 & 0.1333 & 0.2543 & 0.1509 & 0.0729 & 0.0611 & 0.1283 & 0.0803 & 0.1399 & 0.1209 & 0.1831 & 0.1281 & 0.1001 & 0.0871 & 0.1499 & 0.1093 \\
& DVBPR    & 0.1799 & 0.1447 & 0.2688 & 0.1569 & \underline{0.0848} & \underline{0.0639} & \underline{0.1358} & \underline{0.0811} & 0.1427 & 0.1231 & 0.1834 & 0.1283 & 0.1092 & 0.0896 & 0.1577 & 0.1103 \\
& AMR      & 0.1758 & 0.1266 & 0.2681 & 0.1496 & 0.0799 & 0.0578 & 0.1293 & 0.0791 & 0.1426 & 0.1230 & 0.1880 & 0.1301 & 0.1101 & 0.0902 & 0.1573 & 0.1109 \\
\midrule
\multirow{2}{*}{Multimodal}
& MMGCN    & 0.1753 & 0.1201 & 0.2591 & 0.1544 & 0.0841 & 0.0600 & 0.1286 & 0.0793 & 0.1430 & 0.1233 & 0.1881 & 0.1344 & 0.1101 & 0.0907 & 0.1564 & 0.1114 \\
& FREEDOM  & 0.1755 & 0.1299 & 0.2603 & 0.1548 & 0.0844 & 0.0610 & 0.1294 & 0.0801 & \underline{0.1432} & \textbf{0.1237} & \underline{0.1882} & \underline{0.1379} & 0.1105 & 0.0910 & \underline{0.1579} & 0.1122 \\
\midrule
\multirow{4}{*}{\shortstack[l]{Multimodal\\Sequential}}
& UniSRec  & 0.1694 & 0.1402 & 0.2511 & 0.1511 & 0.0685 & 0.0603 & 0.1207 & 0.0764 & 0.1330 & 0.1198 & 0.1734 & 0.1354 & 0.1045 & 0.0833 & 0.1403 & 0.1089 \\
& MP4SR   & 0.1783 & 0.1493 & 0.2664 & 0.1534 & 0.0748 & 0.0610 & 0.1222 & 0.0810 & 0.1293 & 0.1142 & 0.1684 & 0.1322 & 0.1067 & 0.0849 & 0.1426 & 0.1103 \\
& MMSR     & 0.1780 & \underline{0.1578} & 0.2683 & 0.1604 & 0.0640 & 0.0583 & 0.1172 & 0.0734 & 0.1284 & 0.1203 & 0.1760 & 0.1365 & 0.1107 & 0.0911 & 0.1566 & 0.1209 \\
& HM4SR    & \underline{0.1910} & 0.1545 & \underline{0.2777} & \underline{0.1629} & 0.0633 & 0.0572 & 0.1170 & 0.0712 & 0.1302 & 0.1232 & 0.1863 & \underline{0.1400} & \underline{0.1148} & \underline{0.1004} & 0.1577 & \textbf{0.1225} \\
\midrule
Ours & VisGate  & \textbf{0.2207}$^\dagger$ & \textbf{0.1801}$^\dagger$ & \textbf{0.2889}$^\dagger$ & \textbf{0.1748}$^\dagger$ & \textbf{0.1011}$^\dagger$ & \textbf{0.0759}$^\dagger$ & \textbf{0.1462}$^\dagger$ & \textbf{0.0996}$^\dagger$ & \textbf{0.1449} & \underline{0.1236} & \textbf{0.1907} & \textbf{0.1413} & \textbf{0.1210}$^\dagger$ & \textbf{0.1013} & \textbf{0.1681}$^\dagger$ & \underline{0.1224} \\
\midrule
\multicolumn{2}{l|}{\% Improve} & +15.5 & +13.5 & +7.1 & +7.3 & +19.2 & +18.8 & +7.7 & +22.8 & +1.2 & -- & +1.0 & +2.5 & +3.9 & +0.8 & +6.5 & -- \\
\bottomrule
\end{tabular}
\label{tab:main_results}
\end{table*}

\subsection{RQ1: Is Visual Utility Item-Dependent?}
\begin{figure}[!t]
\centering
\includegraphics[trim=0 4.5pt 0 0, clip, width=0.35\textwidth]{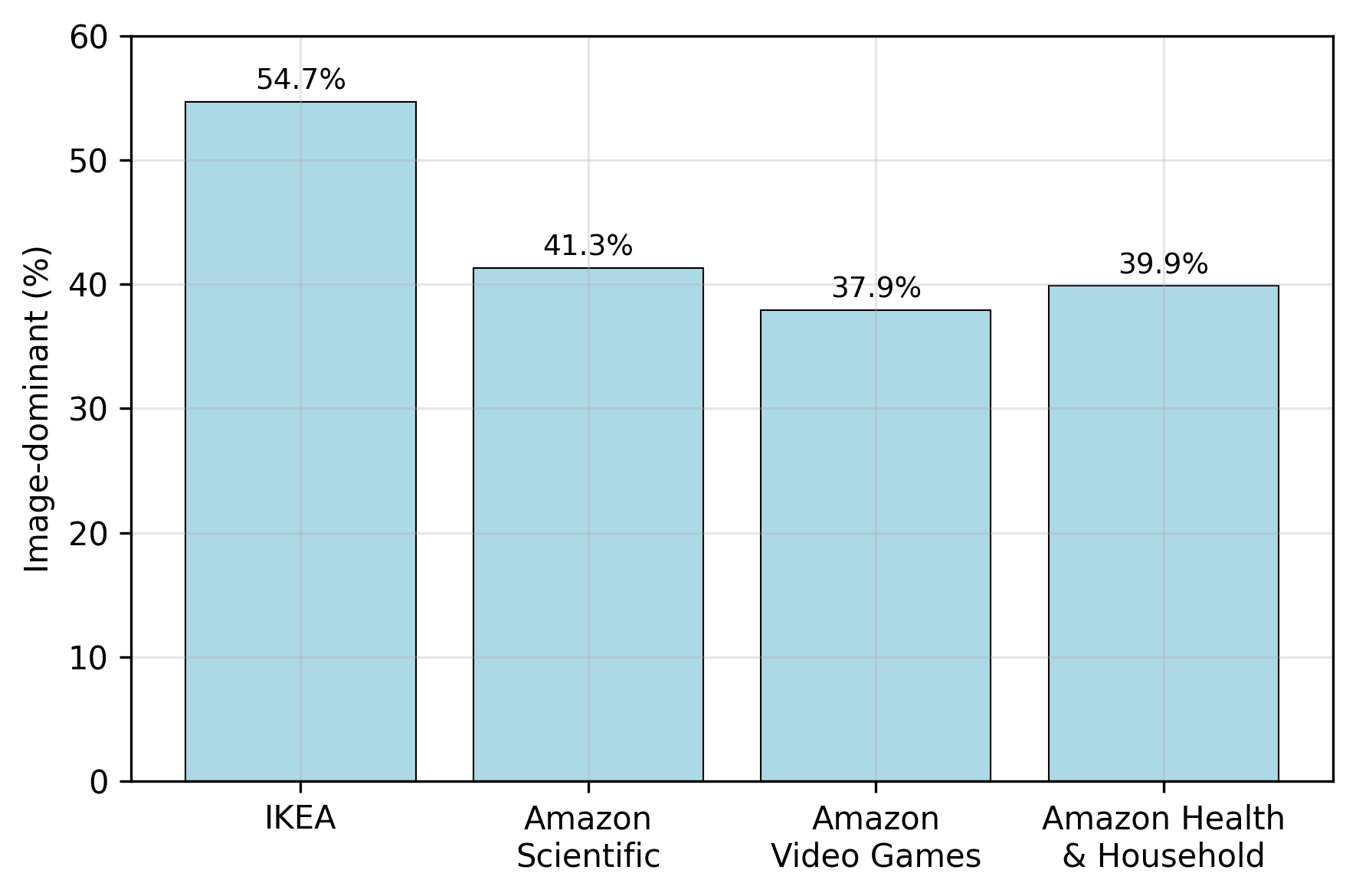} 
\caption{Percentage of images used for the datasets based on the item-level gating function.}
\label{fig:embedding_distribution}
\end{figure}

\begin{figure}[H]
    \centering
    \includegraphics[width=0.45\linewidth,clip]{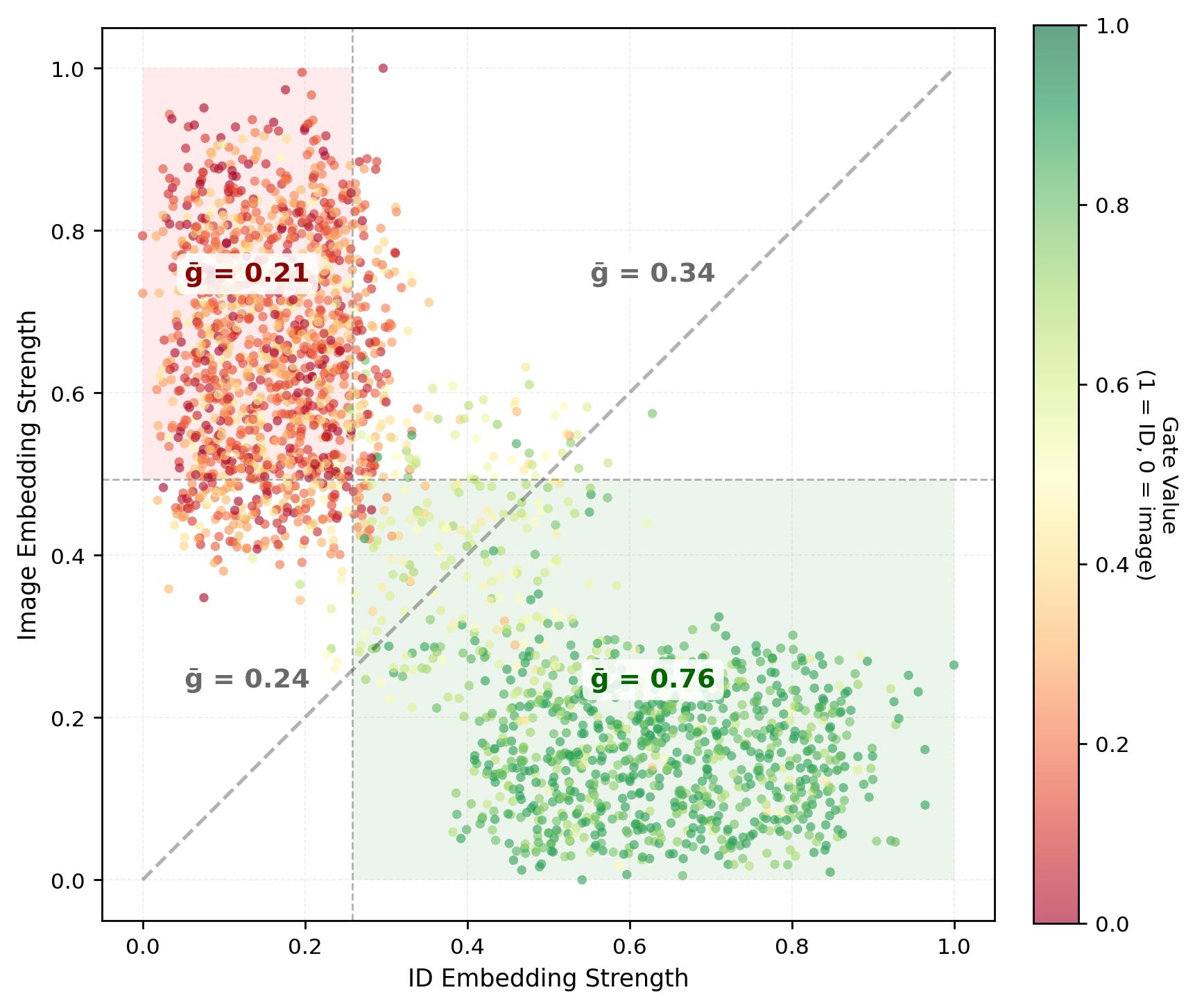}
    \hfill
    \includegraphics[width=0.45\linewidth,clip]{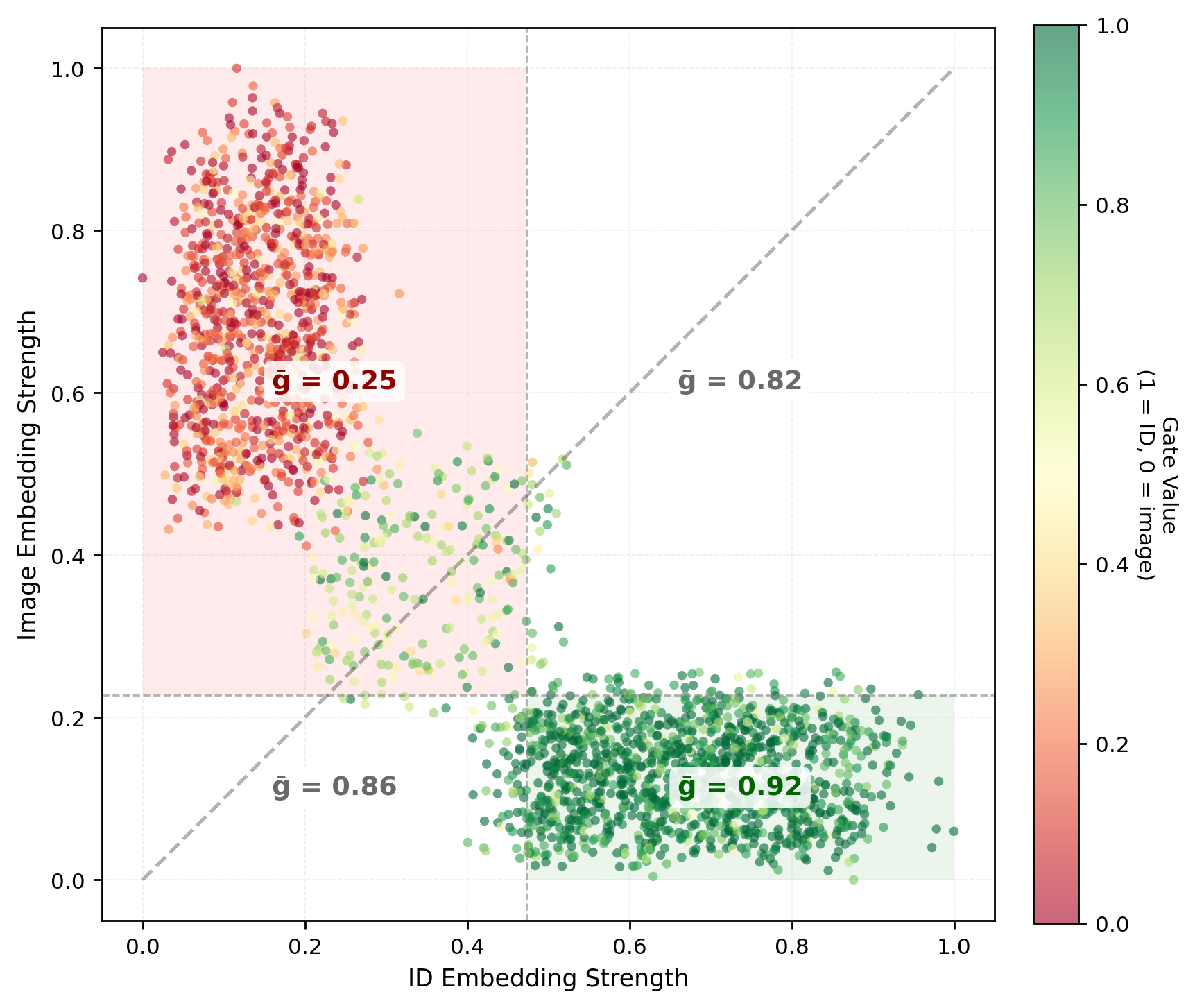}
    \\
    \textbf{(a) IKEA \hspace{3.5cm} (b) Amazon H\&H}
    \caption{Gate value $g_i$ as a function of ID and image embedding strength($\ell_2$ norm, min-max normalised). Dashed lines mark the per-axis medians, partitioning items into four equal-population quadrants. Mean gate values$\bar{g}$ are annotated in each quadrant; the shaded regions (red: strong-image/ weak-ID; green: strong-ID / weak-image) highlight the two complementary quadrants where the gate difference is largest.}
    \label{fig:learn_gate}
\end{figure}

From Table \ref{tab:gating_results} we can see that item-level gating performs better than fixed gate values, confirming that no single modality weight suits all items. Additionally, from Figure \ref{fig:embedding_distribution} we identify that within each dataset, the distribution of learned gate values is broad rather than peaked, showing that the gate makes genuinely item-level decisions. The quadrant analysis in Figure \ref{fig:learn_gate} further reveals a complementary pattern where items with strong image but weak ID embeddings receive low gate values ($\bar{g} = 0.21$ on IKEA), while strong-ID/weak-image items receive high ones ($\bar{g} = 0.76$), with gaps of ${\approx}0.55$--$0.67$ between complementary quadrants across datasets.

\begin{table}[h!]
\centering
\small
\caption{Uniform fusion at fixed gate values vs. learned item-level gating.}
\begin{tabular}{lcccc}
\hline
 & \multicolumn{3}{c}{Uniform Gating} & \\
\cline{2-4}
Dataset & 0.3 & 0.5 & 0.7 & VisGate \\
\hline
IKEA (H@10)        & 0.2619 & 0.2795 & 0.2636 & \textbf{0.2889} \\
Amazon H\&H (H@10) & 0.1577 & 0.1492 & 0.1309 & \textbf{0.1681} \\
\hline
\end{tabular}
\label{tab:gating_results}
\end{table}

\subsection{RQ2: Does Visual Utility Increase Under Sparsity?}
The largest performance gains appear on the two sparsest datasets, IKEA and Amazon Scientific, where ID embeddings are least reliable. At the item level, weak-ID items consistently receive lower gate values, while strong-ID items rely more on collaborative signals. The temporal analysis in Table \ref{tab:temporal_gate} confirms this dynamically. We see that early interactions receive lower gate values than recent ones across all datasets, with the effect most pronounced on IKEA and Amazon Scientific, where collaborative signal accumulates more slowly.

\begin{table}[t]
\centering
\small
\caption{Mean gate value $\bar{g}$ for early (first half) vs.\ recent 
(second half) interactions by sequence position, averaged across users. 
Lower values indicate greater visual reliance.}
\label{tab:temporal_gate}
\setlength{\tabcolsep}{5pt}
\begin{tabular}{l c c c}
\toprule
Dataset & Early $\bar{g}$ & Recent $\bar{g}$ & $\Delta$ \\
\midrule
IKEA                       & 0.441 & 0.468 & +0.027 \\
Amazon Scientific          & 0.573 & 0.601 & +0.028 \\
Amazon Video Games         & 0.607 & 0.621 & +0.014 \\
Amazon Health \& Household & 0.588 & 0.603 & +0.015 \\
\bottomrule
\end{tabular}
\end{table}

\subsection{RQ3: Does Visual Utility Correlate With Visual Distinctiveness?}
\begin{figure}[h!]
\centering
\includegraphics[trim=0 15mm 0 20mm, clip, width=0.5\textwidth]{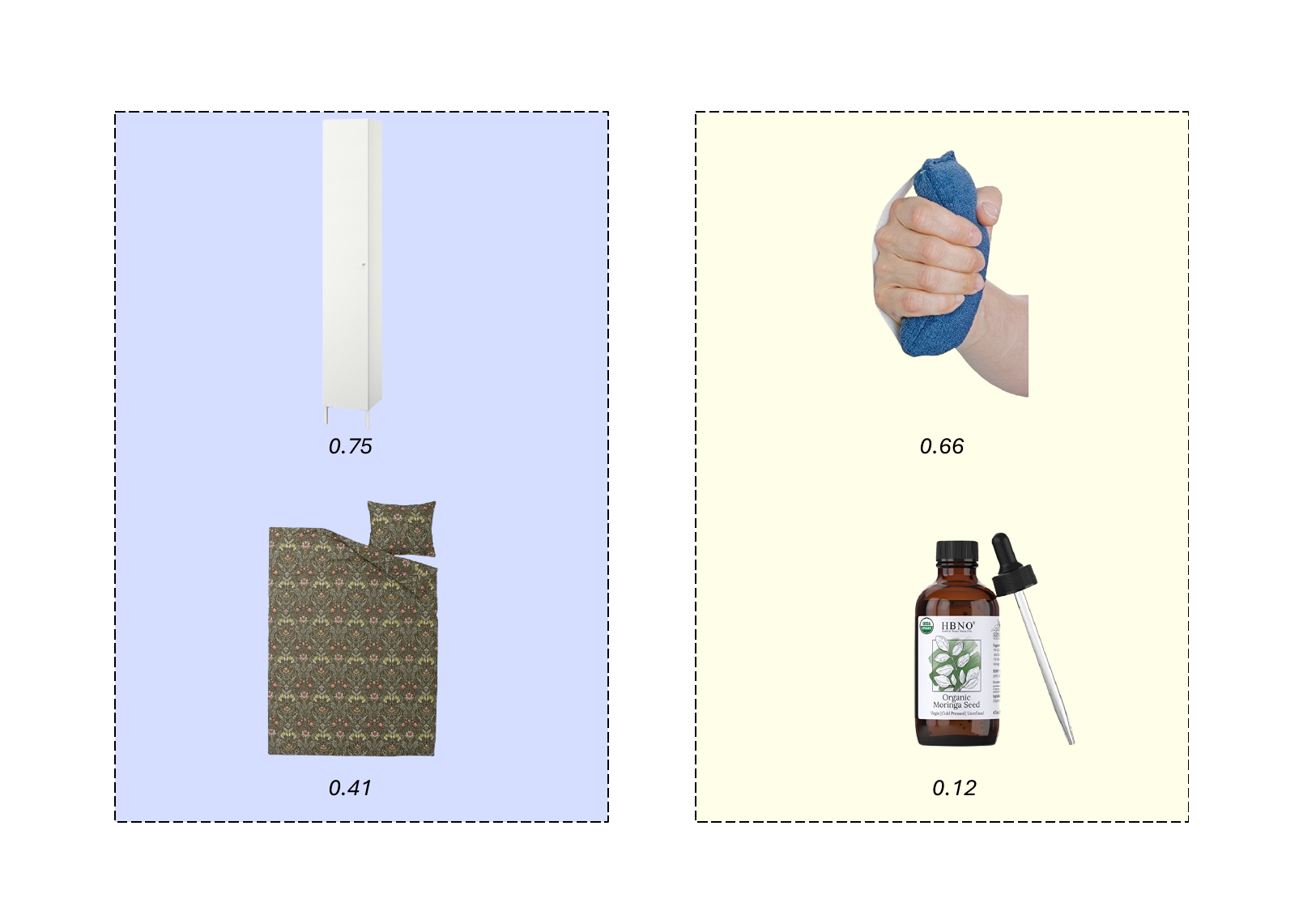}
\caption{Pictures of images with highest gate (top) and lowest (bottom) based on their  average gating values with IKEA on the left hand side and Amazon H\&H on the right hand side.}
\label{fig:gate_qual}
\end{figure}

Figure \ref{fig:gate_qual} shows items at the extremes of the gate distribution. Low-gated items are visually distinctive. For example, patterned bed sheets or products with clear branding, where appearance directly influences purchase decisions \cite{shukla2022influence, gupta2025psychology, orquin2020visual}. High-gated items have generic appearances, such as a plain white bookself. Here, interaction patterns are more informative. This confirms the gate assigns visual weight in semantically meaningful ways consistent with consumer behaviour research.

\begin{figure}[h]
    \centering
    \includegraphics[trim=0 15mm 0 20mm, clip, width=0.5\textwidth]{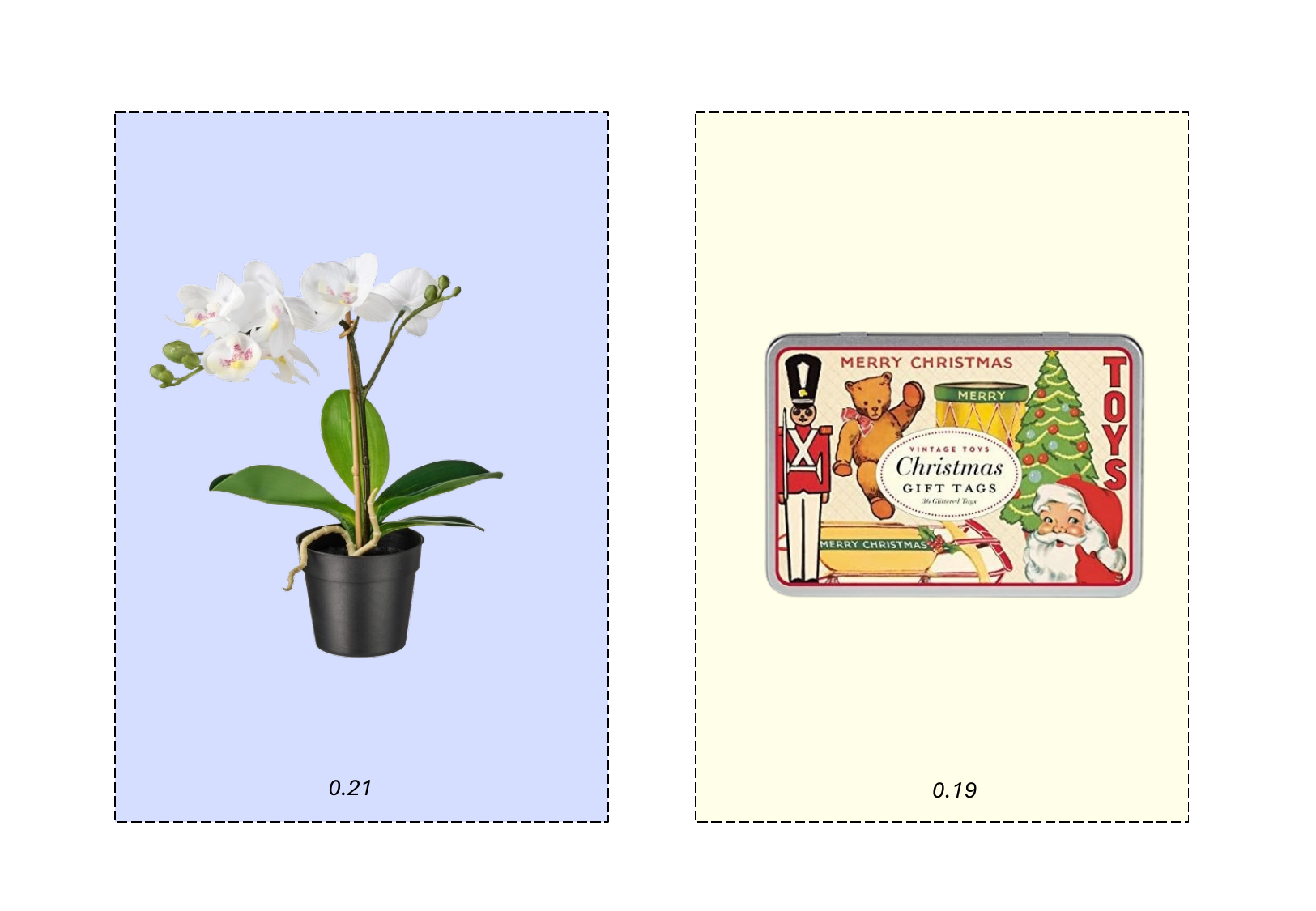}
    \caption{Failure cases from IKEA (left) and Amazon Health \& 
    Household (right). Items have visually distinctive appearances 
    and low gate values (high visual reliance) but few interactions, 
    leading to poorly calibrated visual representations. Gate values 
    and mean interaction counts are shown below each item.}
    \label{fig:failure}
\end{figure}

However, the gate is an imperfect proxy. The most consistent failure occurs with visually distinctive but interactionally sparse items as seen in Figure \ref{fig:failure}. The gate correctly identifies weak collaborative signal but incorrectly infers strong visual signal, when both are unreliable due to limited co-occurrence supervision. Table \ref{tab:failure} quantifies this. The failure rate is highest in the low-gate, low-interaction quadrant. A simple interaction-count threshold could serve as a post-hoc correction, and should be explored in future work. 

\begin{table}[t]
\centering
\small
\caption{Fraction of items where removing visual features improves 
Hit@10, partitioned by gate value and interaction count (IKEA). 
Low gate = high visual reliance.}
\label{tab:failure}
\setlength{\tabcolsep}{4pt}
\begin{tabular}{l c c}
\toprule
 & Low interaction count & High interaction count \\
\midrule
Low gate (visual) & 0.31 & 0.09 \\
High gate (collab) & 0.07 & 0.04 \\
\bottomrule
\end{tabular}
\end{table}

\subsection{Ablation Study}
Table \ref{tab:ablation_results} summarises ablation results. Removing the contrastive loss has the largest impact, confirming that complementarity-preserving supervision is the most important component. Removing gating causes the second-largest drop. To better understand the effect of the sequence-context of the gate, we compare to an item-only variant that does not include the sequence context. This leads to a drop in performance, confirming that visual utility depends on user state as well as item properties. Joint end-to-end training performs worse than the two-stage design, suggesting joint optimisation causes the projector to drift from co-occurrence grounded representations.

\begin{table}[t]
\centering
\small
\caption{Ablation study on IKEA and Amazon Health \& Household datasets.}
\label{tab:ablation_results}
\setlength{\tabcolsep}{4pt}
\begin{tabular}{l l c c}
\toprule
Dataset & Variant & Hit@10 & NDCG@10 \\
\midrule
\multirow{6}{*}{IKEA} 
& w/o Gating                     & 0.2711 & 0.1669 \\
& w/o Co-occurrence Learning     & 0.2658 & 0.1629 \\
& w/o Contrastive Loss           & 0.2311 & 0.1572 \\
& Joint End-to-End Training      & 0.2743 & 0.1673 \\
& w/o Sequence Context           & 0.2799 & 0.1711 \\
& \textbf{VisGate (full)}        & \textbf{0.2889} & \textbf{0.1748} \\
\midrule
\multirow{6}{*}{Amazon H\&H} 
& w/o Gating                     & 0.1492 & 0.1034 \\
& w/o Co-occurrence Learning     & 0.1501 & 0.1109 \\
& w/o Contrastive Loss           & 0.1304 & 0.1011 \\
& Joint End-to-End Training      & 0.1649 & 0.1198 \\
& w/o Sequence Context           & 0.1632 & 0.1203 \\
& \textbf{VisGate (full)}        & \textbf{0.1681} & \textbf{0.1224} \\
\bottomrule
\end{tabular}
\end{table}

\section{Discussion}
These findings propose several principles for context-aware multimodal recommendation that extend beyond sequential systems.

\textbf{Context determines fusion granularity} Item-level gating achieves competitive performance with existing multimodal sequential methods, while consistently outperforming uniform and sequence-level fusion strategies. This suggests that matching fusion decisions to the level at which modality 
informativeness varies is at least as important as the choice of fusion mechanism itself. Modality utility is not a static feature of an item but a latent contextual variable shaped by interaction history and user state.

\textbf{Complementarity should be preserved, not eliminated} The alignment paradigm dominant in multimodal learning optimises for cross-modal retrieval, finding images that match text queries. But for systems that infer context from multiple signals rather than translate between them, alignment reduces the value of fusion by making modalities redundant. Context-aware recommendation stands to benefit most when modalities contribute orthogonal information, with visual, behavioural, and textual signals each capturing dimensions of user intent that the others cannot.

\textbf{Latent context is measurable} Using a learned gate as a proxy for per-item modality contribution is straightforward to implement and produces interpretable estimates of a quantity that is otherwise difficult to observe directly. This has implications beyond visual recommendation. A similar approach 
could support context diagnostics in any setting where the relative informativeness of available signals varies across items, users, or interaction states.

\section{Conclusion and Future Work}\label{future}
This paper investigated when and why visual content provides complementary value to collaborative signals in sequential recommendation, framing the problem as one of latent context modelling. We introduced modality utility as a latent contextual variable and proposed VisGate as both a competitive recommender and a measurement tool for studying it empirically.

Using the learned gate as an analytical lens, we showed that visual utility is item-dependent rather than uniform, increases under interaction sparsity when collaborative signals are weakest, and correlates with visual distinctiveness in semantically meaningful ways. We also identified a systematic failure mode in 
which the gate misfires for visually distinctive but interactionally sparse items, where neither modality is well calibrated. These findings validate the importance of item-level fusion granularity and complementarity preservation. These principles are not specific to our architecture and generalise to any multimodal system where contextual signals vary in informativeness across items or interaction states.

For future work, an open theoretical question is whether the optimal fusion granularity (item, step, sequence) can be predicted from dataset statistics alone, without training a gate. We also plan to extend the framework to richer contextual signals, incorporating textual modalities via LLMs and conditioning the gate on broader sequence-level context, enabling it to capture not only which items benefit from visual signals but when in a user's interaction history those signals are most informative.
\begin{acks}
This work was supported by the Wallenberg AI, Autonomous Systems and Software Program (WASP, http://wasp-sweden.org), funded by the Knut and Alice Wallenberg Foundation.
\end{acks}

\bibliographystyle{ACM-Reference-Format}
\bibliography{sample-base}

\end{document}